\documentclass{article}

\title{A Note on Pointwise Dimensions}
\author{
	Neil Lutz\footnote{Research supported in part by National Science Foundation Grant 1445755.}\\
	\\Department of Computer Science, Rutgers University
	\\Piscataway, NJ 08854, USA
	\\\texttt{njlutz@rutgers.edu}
}

\usepackage{amsmath}
\usepackage{amssymb}
\usepackage{mdframed}
\usepackage{amsthm}
\usepackage{mathtools}
\usepackage{enumitem}
\usepackage{placeins}
\usepackage{url}

\newtheorem{thm}{Theorem}
\newtheorem{obs}[thm]{Observation}

\newtheorem*{defn}{Definition}

\DeclareMathOperator{\Dim}{Dim}

\newcommand{\R}{\mathbb{R}}

\newcommand{\N}{\mathbb{N}}
\newcommand{\Q}{\mathbb{Q}}

\begin{document}
	\maketitle
	\begin{abstract}
		This short note describes a connection between algorithmic dimensions of individual points and classical pointwise dimensions of measures.
	\end{abstract}
	\section{Introduction}
	Effective dimensions are pointwise notions of dimension that quantify the density of algorithmic information in individual points in continuous domains. This note aims to clarify their relationship to the classical notion of pointwise dimensions of measures, which is central to the study of fractals and dynamics. This connection is then used to compare algorithmic and classical characterizations of Hausdorff and packing dimensions. See~\cite{Mayo08,Reim04} for surveys of the strong ties between algorithmic information and fractal dimensions.

	\section{Pointwise Notions of Dimension}
	We begin by defining the two most well-studied formulations of algorithmic dimension~\cite{DowHir10}, the effective Hausdorff and packing dimensions. Given a point $x\in\R^n$, a precision parameter $r\in\N$, and an oracle set $A\subseteq\N$, let
	\[K^A_r(x)=\min\{K^A(q):q\in\Q^n\cap B_{2^{-r}}(x)\}\,,\]
	where $K^A(q)$ is the prefix-free Kolmogorov complexity of $q$ relative to the oracle $A$, as defined in~\cite{LiVit08}, and $B_{2^{-r}}(x)$ is the closed ball of radius $2^{-r}$ around $x$.
	\begin{defn}
		The \emph{effective Hausdorff dimension} and \emph{effective packing dimension} of $x\in\R^n$ relative to $A$ are
		\begin{align*}
			\dim^A(x)&=\liminf_{r\to\infty}\frac{K^A_r(x)}{r}\\
			\Dim^A(x)&=\limsup_{r\to\infty}\frac{K^A_r(x)}{r}\,,
		\end{align*}
		respectively~\cite{Lutz03b,Mayo02,AHLM07}.
	\end{defn}
	Intuitively, these quantities are the (lower and upper asymptotic) density of algorithmic information in $x$.
	In the classical setting, pointwise dimensions are defined for a given measure according to its (lower and upper asymptotic) rate of decay around $x$.
	\begin{defn}
		For any locally finite measure $\mu$ on $\R^n$, the \emph{lower} and \emph{upper pointwise dimension} of $\mu$ at $x\in\R^n$ are
	\begin{align*}
		\dim_\mu(x)&=\liminf_{\rho\to 0}\frac{\log\mu(B_{\rho}(x))}{\log \rho}\\
		\Dim_\mu(x)&=\limsup_{\rho\to 0}\frac{\log\mu(B_{\rho}(x))}{\log \rho}\,,
	\end{align*}
	respectively~\cite{Falc97}.
	\end{defn}
	As Young~\cite{Youn82} notes, these limits are unchanged if $\rho$ is replaced by any sequence $\{\rho_r\}_{r\in\N}$ satisfying $\rho_r\downarrow 0$ and $\log \rho_{r+1}/\log\rho_r\to 1$. In particular, the sequence $\{2^{-r}\}_{r\in\N}$ may be used. Also, this definition in no way relies on additivity, so it applies equally well to outer measures and semimeasures.
	
	We relate these two notions of pointwise dimension by defining an outer measure $\kappa^A$ on $\R^n$ for any given oracle set $A\subseteq\N$. For every $E\subseteq\R^n$,
	\[\kappa^A(E)=2^{-K^A(E)}\,,\]
	where, following Shen and Vereschagin~\cite{SheVer02},
	\[K^A(E)=\min_{q\in E\cap\Q^n}K^A(q)\,.\]
	This minimum is taken to be infinite when $E\cap\Q^n=\emptyset$. It is easy to see that $\kappa^A$ is also subadditive and monotonic, and that $\kappa^A(\emptyset)=0$. Since $K^A$ is non-negative, $\kappa^A$ is finite.
	\begin{obs}\label{obs:kappa}
		For every oracle set $A\subseteq\N$ and all $x\in\R^n$,
		\begin{align*}
			\dim_{\kappa^A}(x)&
			=\dim^A(x)\,.\\
			\Dim_{\kappa^A}(x)&
			=\Dim^A(x)\,.
		\end{align*}
	\end{obs}

	This fact is closely related to (and was observed independently of) an unpublished remark by Reimann~\cite{Reim14} stating that $\dim(x)$ is equal to the pointwise dimension at $x$ of Levin's~\cite{Levi74} universal lower semicomputable continuous semimeasure.
	
	Pointwise dimensions of measures give rise to global dimensions of measures, which we now briefly comment on. In classical fractal geometry, the global dimensions of Borel measures play a substantial role in studying the interplay between local and global properties of fractal sets and measures.
	
	\begin{defn}
		For any locally finite Borel measure $\mu$ on $\R^n$ and $x\in\R^n$, the \emph{lower} and \emph{upper Hausdorff and packing dimension} of $\mu$ are
		\begin{align*}
		\dim_H(\mu)&=\sup\{\alpha : \mu(\{x : \dim_\mu(x) < \alpha\}) = 0\}
		\\\Dim_H(\mu)&=\inf\{\alpha : \mu(\{x : \dim_\mu(x) > \alpha\}) = 0\}
		\\\dim_P(\mu)&=\sup\{\alpha : \mu(\{x : \Dim_\mu(x) < \alpha\}) = 0\}
		\\\Dim_P(\mu)&=\inf\{\alpha : \mu(\{x : \Dim_\mu(x) > \alpha\}) = 0\}\,,
		\end{align*}
		respectively~\cite{Falc97}.
	\end{defn}
	Extending these definitions to outer measures, we may consider global dimensions of the outer measures $\kappa^A$. For every $A\subseteq\N$, $\kappa^A$ is supported on $\Q^n$ and $\dim^A(p)=0$ for all $p\in\Q^n$, which implies the following.
	\begin{obs}\label{obs:globaldim}
		For every $A\subseteq\N$,
		\[\dim_H(\kappa^A)=\Dim_H(\kappa^A)=\dim_P(\kappa^A)=\Dim_P(\kappa^A)=0\,.\]
	\end{obs}

	\section{Pointwise Principles for Dimensions of Sets}
	The \emph{point-to-set principle} of J. Lutz and N. Lutz expresses classical Hausdorff and packing dimensions in terms of relativized effective Hausdorff and packing dimensions.
	\begin{thm}[J. Lutz and N. Lutz~\cite{LutLut17}]\label{thm:p2s}
		For every nonempty $E\subseteq\R^n$,
		\begin{enumerate}
			\item[\textup{(a)}] $\displaystyle \dim_H(E)=\adjustlimits\min_{A\subseteq\N}\sup_{{x}\in E}\,\dim^A({x})\,.$
			\item[\textup{(b)}] $\displaystyle \dim_P(E)=\adjustlimits\min_{A\subseteq\N}\sup_{{x}\in E}\,\Dim^A({x})\,.$
		\end{enumerate}
	\end{thm}
	In light of Observation~\ref{obs:kappa}, this principle may be considered a member of the family of results, such as Billingsley's lemma~\cite{Bill61} and Frostman's lemma~\cite{Fros35}, that relate the local decay of measures to global properties of measure and dimension. Useful references on such results include~\cite{BisPer17,Hoch14,Matt95}.
	
	Among classical results, this principle is most directly comparable to 
	the \emph{weak duality principle} of Cutler~\cite{Cutl94} (see also~\cite{Falc97}),
	which expresses Hausdorff and packing dimensions in terms of lower and upper pointwise dimensions of measures. For nonempty $E\subseteq\R^n$, let $\mathcal{P}(E)$ be the collection of Borel probability measures on $\R^n$ such that the $E$ is measurable and has measure 1, and let $\overline{E}$ be the closure of $E$.
	\begin{thm}[Cutler~\cite{Cutl94}] For every nonempty $E\subseteq\R^n$,
		\begin{enumerate}
			\item[\textup{(a)}] $\displaystyle \dim_H(E)=\adjustlimits\inf_{\mu\in\mathcal{P}(\overline{E})}\sup_{{x}\in E}\,\dim_\mu({x})\,.$
			\item[\textup{(b)}] $\displaystyle \dim_P(E)=\adjustlimits\inf_{\mu\in\mathcal{P}(\overline{E})}\sup_{{x}\in E}\,\Dim_\mu({x})\,.$
		\end{enumerate}
	\end{thm}
	By letting $\mathcal{A}=\{\kappa^A:A\subseteq\N\}$, Theorem~\ref{thm:p2s} can be restated even more similarly as $\dim_H(E)=\inf_{\mu\in\mathcal{A}}\sup_{x\in E}\dim_\mu(x)$ and $\dim_P(E)=\inf_{\mu\in\mathcal{A}}\sup_{x\in E}\Dim_\mu(x)$. Notice, however, that the collections over which the infima are taken in these two results, $\mathcal{A}$ and $\mathcal{P}(\overline{E})$, are disjoint and qualitatively very different. In particular, $\mathcal{A}$ does not depend on $E$. Whereas the global dimensions of the measures in $\mathcal{P}(\overline{E})$ are closely tied to the dimensions of $E$~\cite{Falc97}, Observation~\ref{obs:globaldim} shows that the outer measures in $\mathcal{A}$ all have trivial global dimensions.
\bibliography{npd}
\end{document}